\documentclass[cits]{PoS}

\usepackage{natbib}
\bibpunct{[}{]}{,}{a}{}{;} 

\title{A Global Energetic Model for Microquasars (GEMM): A rich and consistent disk+jet solution.}

\ShortTitle{GEMM}

\author{\speaker{C\'edric Foellmi}\thanks{C.F. acknowledges support from the Swiss National Science Foundation (grant PA002-115328).}, Pierre-Olivier Petrucci\thanks{This work has been supported by the ANR-05-JC42835 project funded by the "Agence National de la Recherche".}, Jonathan Ferreira and Gilles Henri\\
        Laboratoire d'Astrophysique de Grenoble, OSUG, Universit\'e Joseph Fourier, CNRS UMR 5571, 414 Rue de la Piscine, 38400 Saint-Martin d'H\`eres, France.\\
        E-mails: \email{cedric.foellmi@obs.ujf-grenoble.fr}, \email{pierre-olivier.petrucci@obs.ujf-grenoble.fr}, \email{jonathan.ferreira@obs.ujf-grenoble.fr}, \email{gilles.henri@obs.ujf-grenoble.fr}}

\author{}

\abstract{Based on a dynamical model describing how stationary, powerful and self-collimated jets are being launched from a magnetized disk, we build a consistent disk+jet microquasar picture. Our disk is a new type of disk solution called the Jet Emitting Disk (JED), and whose characteristics are directly constrained by the presence of a jet. We assume a one-temperature plasma with thermal particles only. By solving the radiative equilibrium of the disk, we obtain three branches of solutions, a hot and a cold ones (both thermally stable), and an intermediate one, thermally unstable. The hot solution possess the global observed characteristics of what has been often called a "corona" located above the inner disk region. We present this new disk solution, and how the radiative equilibrium is computed. We discuss the richness of the solution, and show the ability of the model to reproduce an observed spectral energy distribution of XTE J1118+480 with reasonable parameters. We finally outline some perspectives of the model.}

\FullConference{VII Microquasar Workshop: Microquasars and Beyond\\
		 September 1-5 2008\\
		 Fo\c{c}a, Izmir, Turkey}

\begin{document}
\section{The ubiquity and origin of jets.}

In the goal of understanding microquasars, their evolution and the physics their harbor, one is faced with great challenges when attempting to reproduce the various kind of observations and the abundant phenomenology accumulated over the years. The most common types of observations are spectral energy distributions from radio to X-rays, the timing properties and variability, and finally the images of compact, collimated jets. Since no definite picture exists so far on the underlying accretion disk structure, the issue of the rapid variability, and in particular the Quasi-Periodic Oscillations (QPOs), is nowadays tackled with dedicated models (e.g. \citet{Tagger-Varniere-2006}, see also \pos{PoS(MQW7)012}). As for the jets, there is a consensus, supported by many observations, that radio emission from microquasars is dominated by the synchrotron radiation from a compact jet \citep[e.g.][]{Heinz-Sunyaev-2003}. Remains the disks, and to find a consistent picture of their evolution, that is certainly more related to the combination of the various components, and the consistency of this combination. Recent reviews (see for instance Gallo et al., in these proceedings: \pos{PoS(MQW7)001}) mention a family of models, each of them having different strengths and weaknesses. However, one major concern we can have with these models is either their total lack of associated jet, or their lack of physical consistency between disk and jets. 

Our approach is different. We start by acknowledging the presence of jets not only in microquasars, but also in binaries with a neutron star (see for instance Migliari, these proceedings: \pos{PoS(MQW7)015}), Active Galactic Nuclei, young stellar objects (YSO, such as T Tauri stars) and even brown dwarfs \citep{Whelan-etal-2005}. The ubiquity of such jets in objects that have accretion disks suggests a disk-driven origin\footnote{As far as we know, no supersonic jets were so far observed from an object having no accretion disk.}. There have been plenty of theoretical works on the topic of jet launching from disks, often supported by powerful MHD simulations (see for instance the review of Fragile et al., in these proceedings: \pos{PoS(MQW7)039}). Among all these works, we propose that one is actually consistently solving the origin of compact, powerful and persistent jets launched from an accretion disk, provided that a large-scale magnetic field close to equipartition and of same polarity everywhere is present \citep{Ferreira-Pelletier-1993a,Ferreira-Pelletier-1993b,Ferreira-Pelletier-1995}. These theoretical works have also been tested with MHD simulations since then \citep[see for instance][]{Casse-Keppens-2002,Zanni-etal-2007}. The hypothesis of a large scale magnetic field is the main difference with the works presented by Fragile (these proceedings), and its origin must certainly be found in the accretion disk formation and history. 

Within this theoretical framework, jet and disk are part of the same structure; the jet actually imposing conditions on the disk itself. The accretion disk is therefore of a particular type, that we call the Jet Emitting Disk (JED), and will be described in more details in Petrucci et al. (2009, but see also \cite{Ferreira-etal-2006}). One central difference between our JED and an Advection Dominated Accretion Flow (ADAF) is that the accretion power from the disk is not advected onto the black-hole, but rather being used to drive a persistent, powerful and self-collimated MHD jet. Although it has not yet been fully explored, we mention that the ultra-relativistic jets observed in microquasars would come, in our picture, from blobs of electron-positron pairs created inside the funnel of the MHD jet, provided that some conditions are met (see \citep{Ferreira-etal-2006} for more details). These pairs would be continuously heated by the MHD jet and collimated by it as well. 

\begin{figure}[!t]
\centering
\includegraphics[width=.6\textwidth]{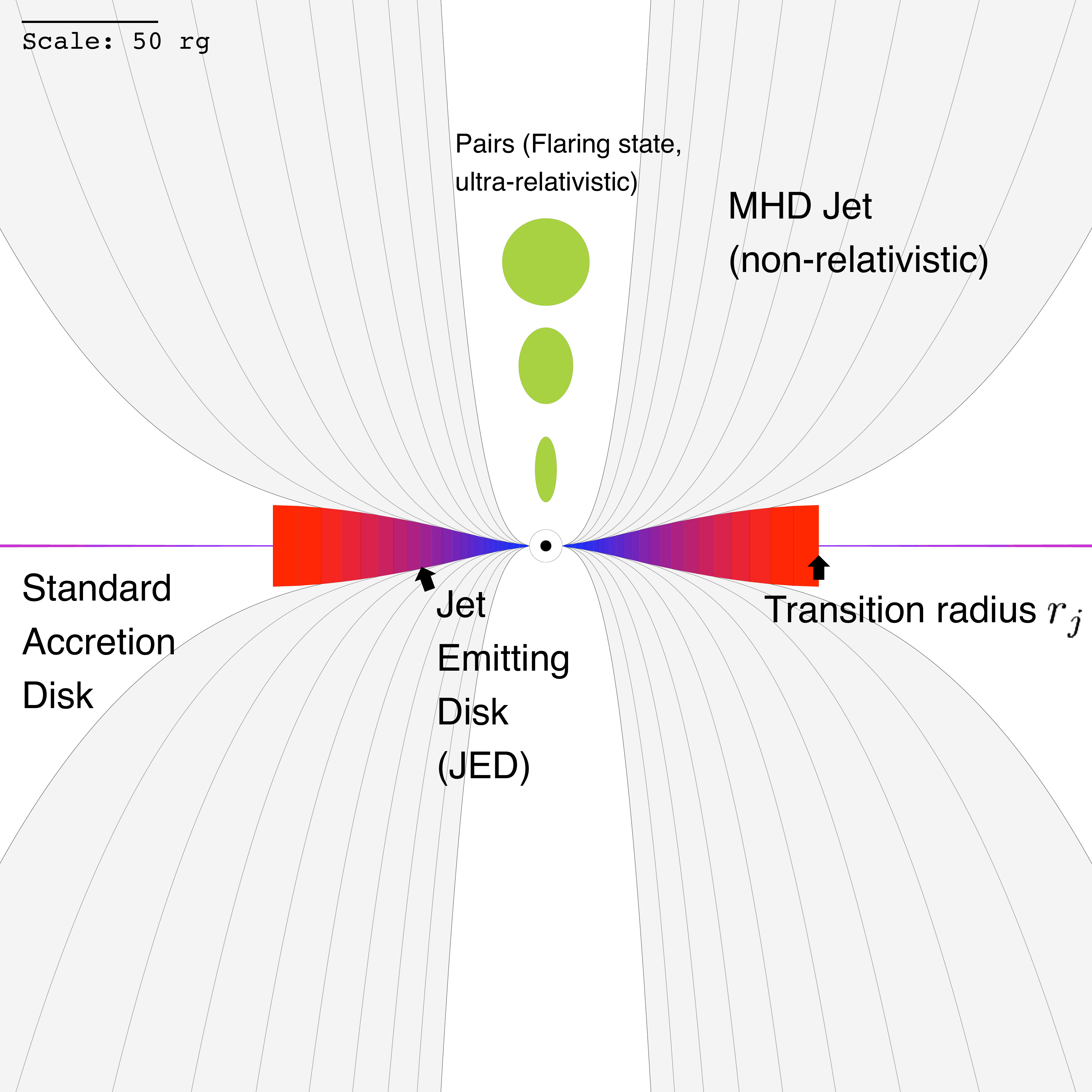} 
\caption{The components of the GEMM: a standard accretion disk in the outer regions transforms itself into a Jet Emitting Disk at transition radius $r_j$. The JED characteristics are imposed by the presence of a steady, powerful, midly-relativistic self-collimated MHD jet. In some special conditions, we expect that pairs are produced inside the MHD funnel, and can produce an ultra-relativistic beam collimated by the MHD jet. The figure is constructed with one of our hot JED solution (see below). The black-hole horizon is drawn at the same scale, and the circle indicates the radius of the last stable orbit $r_{\textrm{\scriptsize{ISCO}}}$. The parameters of the figure are: $M_{bh} = 10\; M_{\odot}$, $\dot{M} = 10^{-1}\; M_{\textrm{\scriptsize{Edd}}} = 10^{-1}\; L_{\textrm{\scriptsize{Edd}}}/c^2$, $r_{j} = 100\;R_{g}$.}
\label{gemm} 
\end{figure} 

The components of our model are sketched in Fig.~\ref{gemm}. The outer disk is a standard accretion disk, down to a transition radius $r_j$, and modeled with a multi-temperature blackbody, as in \citet{Makishima-etal-1986}. It then transforms into a Jet Emitting Disk at $r_j$. The jet is launched from the JED. As it will be shown later, there is no need for a corona since the JED X-ray spectrum reaches 100 keV with thermal particles.

As for now, the transition radius $r_j$ between the standard accretion disk and the JED is a free parameter of our model. We expect that its variations are triggered by the accretion rate and a second parameter, which could be the disk magnetization (see \citet{Petrucci-etal-2008}), defined as: $\mu = B_z^2 / \mu_0 P_{\textrm{{\scriptsize gas}}}$. In our big picture, the magnetization is a function of the disk radius $r$. As we approach the central black hole, $\mu$ increases \citep{Ferreira-etal-2006}. At some transition radius $r_j$ the magnetization goes above a threshold (that should be around unity) the standard accretion disk transforms into a Jet Emitting Disk. The reason for this changes is that the magneto-rotational instability (MRI), which is the phenomenon that is generally accepted as the microphysical source of accretion, does not work with a large value of $\mu$. Therefore, as the magnetic field is advected, we assume that the disk cannot stay as a standard accretion disk, and transits to a JED solution.

\section{The heating and the cooling}

As said above, the jet imposes conditions on the underlying JED. The magnetic torque inside the accretion disk dominates over the viscous (turbulent) torque, and therefore drives the accretion. In a JED solution, the heating of the disk can be inferred from the global energy budget (see the clear introduction on this topic in \citep{Combet-Ferreira-2008}):
\begin{equation}
P_{\textrm{\scriptsize{accretion}}} = P_{\textrm{\scriptsize{jet}}} + P_{\textrm{\scriptsize{rad}}}
\end{equation}
where the accretion power is the difference between the mechanical power entering the JED at the transition radius $r_j$ and goes out at $r_{in}$, the radius of the innermost stable orbit. In a JED the accretion power is shared between the radiative losses of the disk and the power that goes into the jet as MHD Poynting flux. For magnetically driven jets from Keplerian accretion disks, this last term is dominant. Thus, the fraction of energy flux carried away by the jets is written (see \citep{Ferreira-1997}):
\begin{equation}
\frac{P_{\textrm{\scriptsize{jet}}}}{P_{\textrm{\scriptsize{accretion}}}} = \frac{\Lambda}{1+\Lambda} b
\end{equation}
where $\Lambda$ controls the fraction of energy extracted by the large-scale MHD torques (jet) versus that of the small-scale MHD torque (MRI), while $b$ controls the fraction of magnetic energy that feeds the jets versus that being dissipated by the Joule effect within the disk. The parameter $b$ is a free parameter in our model. Note however that self-similar models have shown that $b$ varies between 0.5 and 1.0 (see \citep{Ferreira-1997}). The generic form of the disk heating energy density, vertically integrated, can be written:
\begin{equation}
Q^{+} = \left(1-\frac{\Lambda}{1+\Lambda} b\right) \times \frac{G M \dot{M}}{8 \pi R^3}
\end{equation}
It has been shown (see \citep{Ferreira-2008}) that $\Lambda$ is roughly equal to the inverse disk scale height $\epsilon$: $\Lambda \sim 1/\epsilon$. Assuming $\Lambda \equiv 1/\epsilon$, the JED heating becomes an explicit function of $\epsilon$. On the other hand, since we assume (and a posteriori verify) that the total pressure in the disk is dominated by the gas pressure, the disk vertical equilibrium provides a direct relation between the temperature and the scale height. Consequently, all radiative cooling sources also depends on $\epsilon$. We can therefore compute the radiative equilibrium of the disk: $Q^{-}(\epsilon) = Q_{\textrm{\scriptsize{advection}}} + Q_{\textrm{\scriptsize{radiative}}} = Q^{+}(\epsilon)$, where $Q_{\textrm{\scriptsize{radiative}}} = Q_{\textrm{\scriptsize{Bremsstrahlung}}} + Q_{\textrm{\scriptsize{Synchrotron Compton}}} + Q_{\textrm{\scriptsize{External Compton}}}$.

\section{The Jet Emitting Disk}

In the JED, this equilibrium is computed as follows: we consider a one-temperature plasma, and follow \citet{Esin-etal-1996} for the Bremsstrahlung, and \citet{Mahadevan-etal-1996} for the Synchrotron emission. We include two source of Compton cooling: internal, by the electrons within the JED, and external with photons coming from the standard accretion disk, and reaching the JED. This latter source of cooling appears to be important in some combinations of parameters, and deserve a dedicated geometrical treatment, similarly to what have been done in \citet{Esin-etal-1996}. As for now, we make the approximation that half of the photons produced in the inner radius of the SAD are seen from the JED.

For the Compton cooling, we impose a power-law type of solution, with a Wien bump. To compute the Compton spectrum, we solve the conservations of energy and photons above the cut-off frequency of the input distribution. To obtain the three parameters of the Compton spectrum (namely, the power-law index $\alpha$, the spectrum normalization and the Wien bump normalization $\gamma$) we add a relation between $\alpha$ and $\gamma$ as described in \citet{Wardzinski-Zdziarski-2000} (see their Equ. 23). The mean gain of energy for a Compton scatter is computed with the $\eta$ parameter found in \citet{Dermer-etal-1991}.

\section{My JED solution is rich...}

\begin{figure}[!hb]
\centering
\includegraphics[width=.65\textwidth]{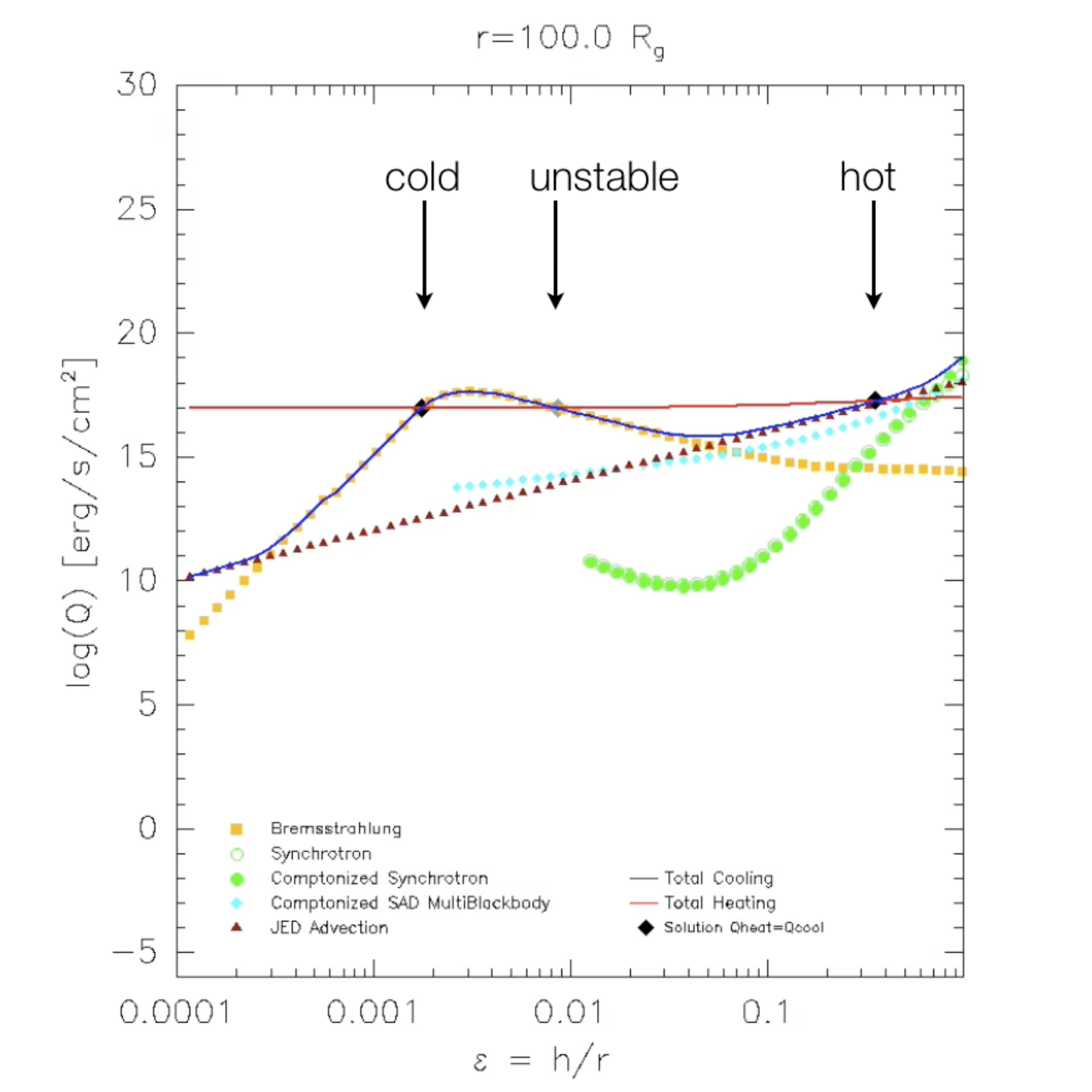}\\
\includegraphics[width=0.5\textwidth]{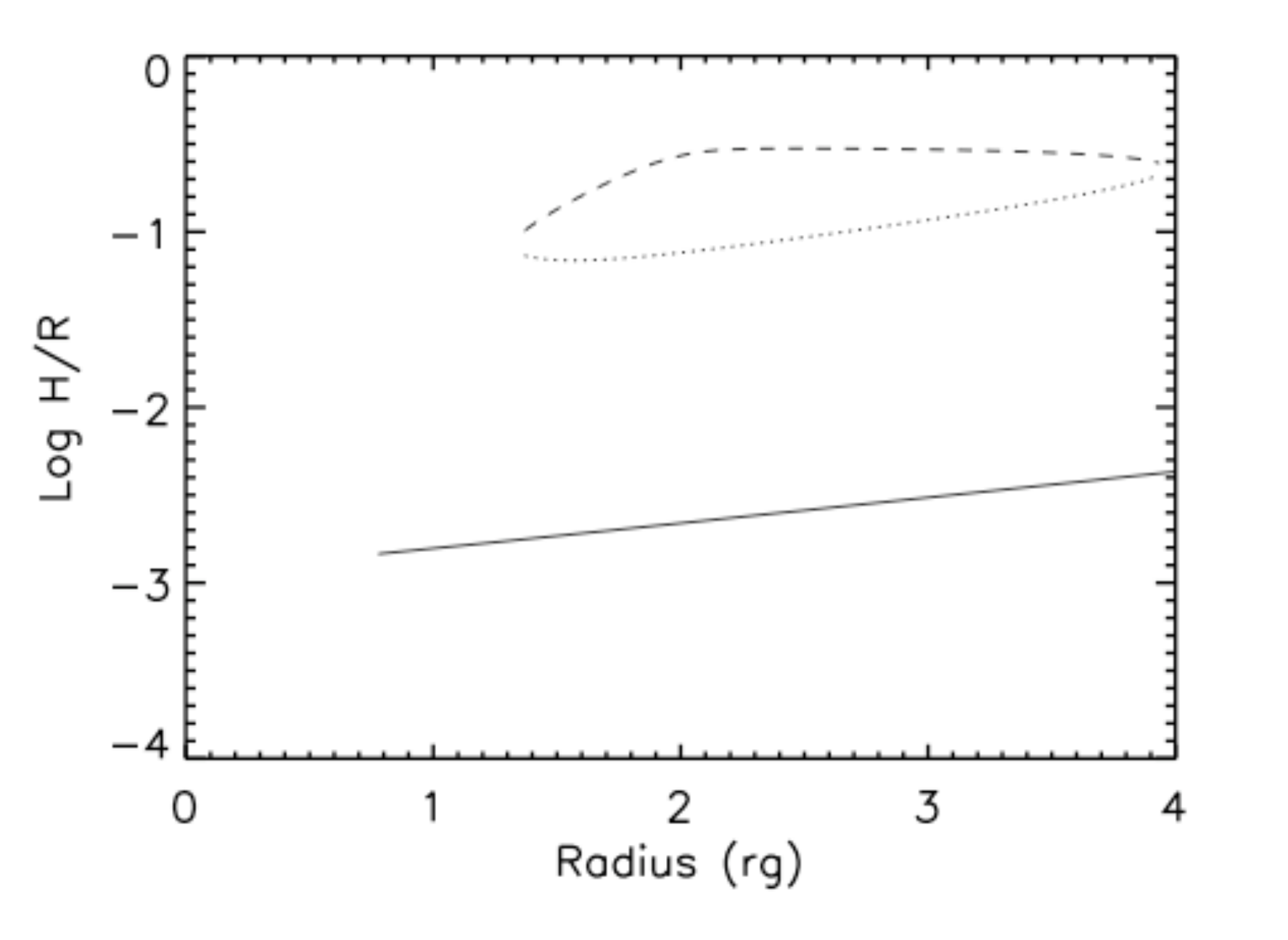}\includegraphics[width=0.5\textwidth]{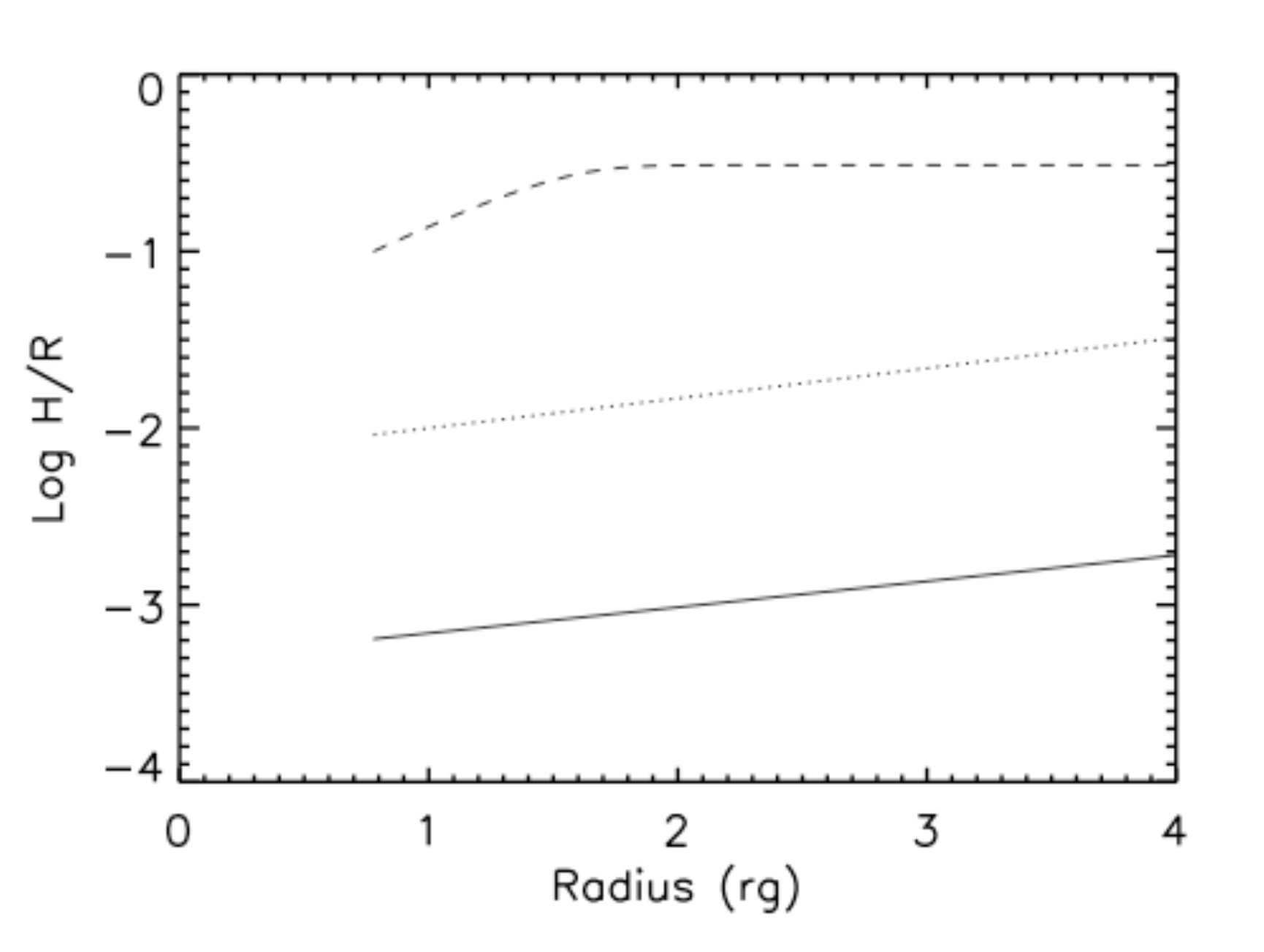} 
\caption{{\bf Top:} The radiative equilibrium is computed here for the given JED radius $r = 100 \;R_{g}$. The figure shows the amount of cooling/heating as function of disk scale height $\epsilon = h/r$ in logarithmic scale. The various cooling contributions are indicated by dots of various shapes. The red line indicate the heating, and the blue line is the total cooling. One can see that the hot disk solution is the result of the interplay between the Comptonized Synchrotron (green dots), the advection (red triangles) and the Compton cooling of SAD photons (in cyan). {\bf Bottom:} Two examples where the three solutions are plotted as function of r, from $r_{\textrm{\scriptsize{ISCO}}} = 6\,R_{g}$ to $r_{j} = 10^{4}\,R_{g}$. Left: $\dot{M} = 10^{-1} M_{\textrm{\scriptsize{Edd}}}$, right: $\dot{M} = 10^{-3} M_{\textrm{\scriptsize{Edd}}}$. The top dashed lines shows the hot solution, while the plain bottom lines indicate the cold ones. We see that in some cases, a solution (here, the hot and unstable ones, in the left left panel) is not present in all radii of the disk. When it happens, we assume that the disk switch rapidly to the other stable one (see Fig.~3).} 
\label{radequ} 
\end{figure} 

\begin{figure}[!hb]
\centering
\includegraphics[width=0.8\textwidth]{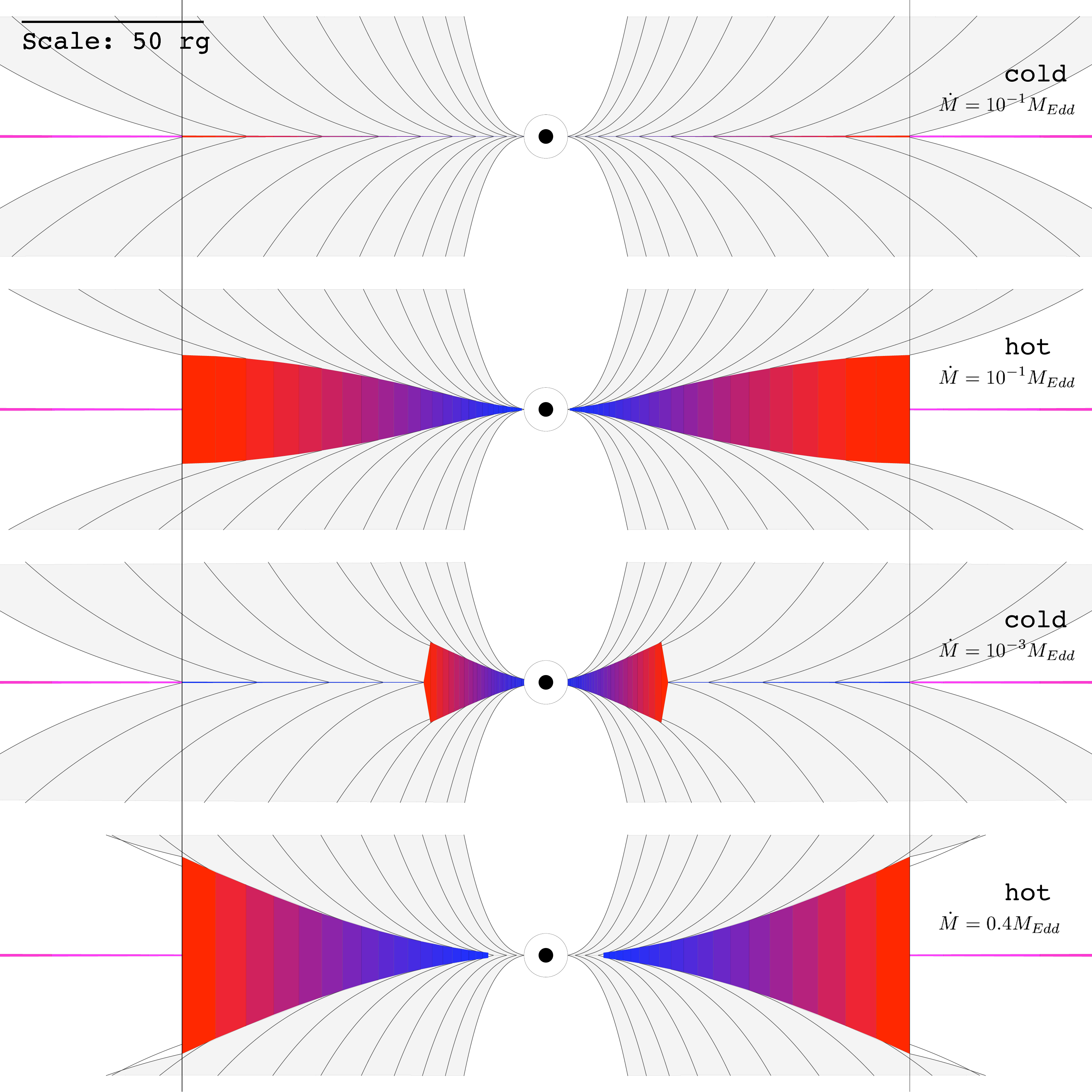} 
\caption{Four different JED solutions resulting from the computation of the radiative equilibrium. The two vertical lines indicate that the position of the transition radius $r_j$ (here 100 $R_g$) is identical in all four solutions in this figure. The first two upper plots show the cold and the hot disk solutions obtained from the exact same input parameters. The next panels illustrate the cases where the hot solution is not present over the full JED extent: in the third panel, the outer JED is cold and the inner JED is hot, while in the fourth one, it is the opposite. The relative scale heights between the disks are respected, and the vertical scale is identical to the horizontal one. The grey areas schematize a jet, to illustrate the fact that a cold JED also launches jets.} 
\label{sols_figs} 
\end{figure} 

The computation of the radiative equilibrium of the disk shows that most often, three solutions exist to the equation $Q^{+} = Q^{-}$: a cold optically thick and a hot optically thin solutions, both thermally stable, bracketing a thermally unstable solution. This is shown in Fig.~\ref{radequ} for a typical example of parameter set. We emphasize here that a cold JED solution is not a standard accretion disk, since it has a (powerful) jet. Moreover, the hot JED solution behaves like a corona, reaching temperature of about 100 keV, and producing the hard X-ray spectrum (see also Petrucci et al. 2009, in prep). For instance, it happens that the hot solution is not always present at all JED radii. It can happen that we have a cold JED in the inner regions, as exemplified in Fig.~\ref{sols_figs}.

\section{First comparison with observations.}

\begin{figure}[!ht]
\centering
\includegraphics[width=0.9\textwidth]{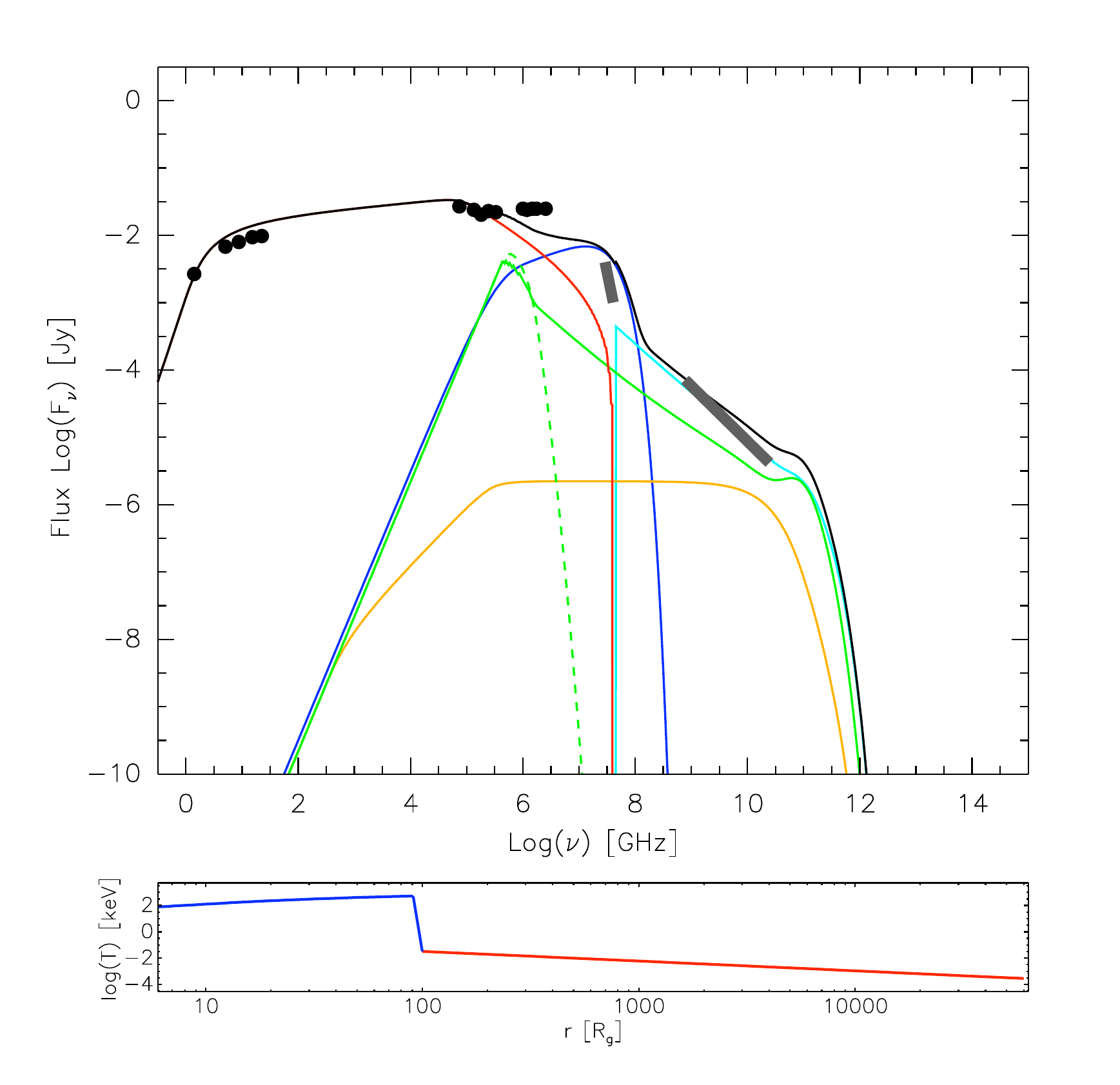} 
\caption{Attempt at reproducing the observed SED of XTE~J1118+480 taken from \citet{Markoff-etal-2001}. The black line indicate our final SED, while the colors indicate the different components: in red the jet, in orange the Bremsstrahlung, in blue the standard accretion disk, in cyan the external Compton. The synchrotron is shown in dashed green as an indication only (it is not counted in the final SED), while the internal Compton on JED synchrotron photons is shown in plain green. The bottom panel shows the corresponding disk where the red line is the standard accretion disk, and in blue the hot JED disk. Parameters are: $M_{bh} = 6. M_{\odot}$, $\dot{M} = 0.2 M_{\textrm{\scriptsize{Edd}}}$ and $D = 1.8$ kpc taken from \citet{Markoff-etal-2001} and a hot JED disk with $r_j = 100 R_g$. The jet emission model follows \citet{Heinz-Sunyaev-2003}, where the synchrotron radiation comes from a distribution of particles described by a power-law: $dn/d\gamma = C \gamma^{-p}$ with $p=2$ and the normalization constant: $C = f \times B^2/(2\mu_0)$. For the fit of XTE~J1118+480, we used $f=0.1$.} 
\label{sed} 
\end{figure}

In order to check our ability to reproduce an observed Spectral Energy Distribution with reasonable parameters, we extracted the data of Fig.~1 of \citet{Markoff-etal-2001} on the galactic microquasar XTE~J1118+480. According to the authors, all points from radio to X-ray were obtained roughly simultaneously. Fig.~\ref{sed} shows that we are able to (manually) reproduce fairly well the observed SED with the following parameters: $M_{bh} = 6 \;M_{\odot}$, $\dot{M} = 0.2 \;M_{\textrm{\scriptsize{Edd}}}$ and $D = 1.8$ kpc taken from \citet{Markoff-etal-2001} and a hot JED disk with $r_j = 100 R_g$.

\section{Perspectives}

We have presented our new disk solution, called the Jet Emitting Disk, and how we calculate spectral energy distributions with it. We have also presented the richness of the radiative equilibrium solutions. Our model, and in particular the way we compute the SEDs and therefore the radiative equilibrium, is far from perfect and still suffer from important approximations. The most critical one is certainly the Comptonization as a power-law in all cases, and its geometrical dependance. In the future, we will test the robustness of our approach of the Compton cooling using for instance the work of Belmont et al. (these proceedings: \pos{PoS(MQW7)040}). 

\begin{figure}[!ht]
\centering
\includegraphics[width=0.8\textwidth]{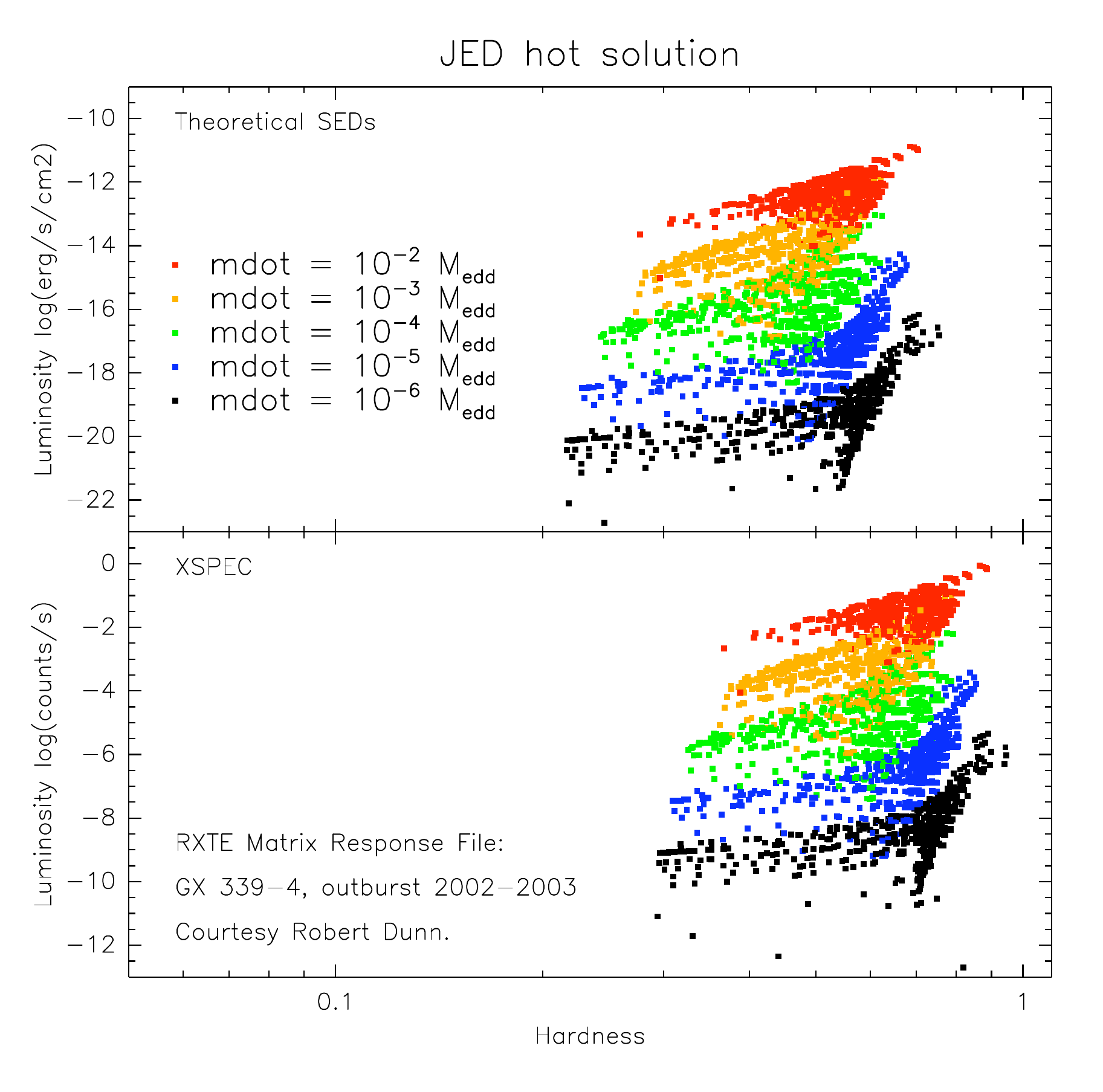} 
\caption{Hardness Intensity Diagrams. We have computed more than 4500 radiative equilibrium to explore the parameter space (here $D = 10$ kpc, $M_{bh} = 10\; M_{\odot}$). We are able to transform our theoretical results into true observable count rates, thanks to RXTE response file kindly provided by Dr. Robert Dunn.}
\label{hid} 
\end{figure}

Our objectives are then not only reproduce with reasonable parameters the observed SEDs in all spectral states, but more importantly explore the Hardness-Intensity Diagram. We have already scratched the surface of this work, and computed a preliminary grid of more than 4500 radiative equilibria, focusing mostly on the low part of the HID, and the low/hard state, as shown in Fig.\ref{hid}. It shows that in the lower-right branch of the hysteresis, the main parameter to increase the luminosity at constant hardness is naturally the accretion rate. Ultimately, we expect to be able to infer the evolution of the main (hopefully non-degenerate) parameters of a microquasar by tracing its hysteresis across the theoretical HID. 

\section*{Acknowledgements}
C.F. acknowledges support from the Swiss National Science Foundation (grant PA002--115328). This work has  extensively used NodeBox (http://nodebox.net), whose authors are hereby acknowledged.
\bibliographystyle{PoS}

\end{document}